\documentstyle[cite,epsfig,sprocl]{article}

\font\eightrm=cmr8

\def\Journal#1#2#3#4{{#1} {\bf #2}, #3 (#4)}

\def\NCA{\em Nuovo Cimento}
\def\NIM{\em Nucl. Instrum. Methods}
\def\NIMA{{\em Nucl. Instrum. Methods} A}
\def\NPB{{\em Nucl. Phys.} B}
\def\PLB{{\em Phys. Lett.}  B}
\def\PRL{\em Phys. Rev. Lett.}
\def\PRD{{\em Phys. Rev.} D}
\def\ZPC{{\em Z. Phys.} C}

\def\st{\scriptstyle}
\def\sst{\scriptscriptstyle}
\def\mco{\multicolumn}
\def\epp{\epsilon^{\prime}}
\def\vep{\varepsilon}
\def\ra{\rightarrow}
\def\ppg{\pi^+\pi^-\gamma}
\def\vp{{\bf p}}
\def\ko{K^0}
\def\kb{\bar{K^0}}
\def\al{\alpha}
\def\ab{\bar{\alpha}}
\def\be{\begin{equation}}
\def\ee{\end{equation}}
\def\bea{\begin{eqnarray}}
 \def\eea{\end{eqnarray}}
\def\CPbar{\hbox{{\rm CP}\hskip-1.80em{/}}}

\begin{document}
 \begin{flushright}
  {\bf Tel-Aviv HEP Preprint} \\
  {\bf TAUP-2598-99} \\
  {\bf 8 September 1999} \\
~~~~~~~~~~~~~~~~~~~~~~~~~~~~ {}\\ 
~~~~~~~~~~~~~~~~~~~~~~~~~~~~ {}\\  
\end{flushright}                                

\title{THE SOURCE SIZE DEPENDENCE ON THE M$_{hadron}$ \\
APPLYING FERMI AND BOSE STATISTICS \\
AND I-SPIN INVARIANCE\,\footnote{Invited talk given by G. Alexander
at the XXIX Int. Symp. on Multiparticle Dynamics,
 9--13 August 1999, Providence RI, USA. (to be published
 in the proceedings of this conference)}}

\author{ \bf  Gideon Alexander and Iuliana Cohen }
\author{  }
\address{School of Physics and Astronomy,
 Tel-Aviv University, Tel-Aviv, Israel}




\maketitle\abstracts{
The 
emission volume sizes of pions and Kaons,
$r_{\pi^{\pm} \pi^{\pm}}$ and $r_{K^{\pm}K^{\pm}}$, 
measured in the hadronic Z$^0$ decays
via the Bose-Einstein Correlations (BEC), and the recent  
measurements of $r_{\Lambda\Lambda}$ obtained by
through the Pauli exclusion principle are used to study
the $r$ dependence on the hadron mass. A clear
\mbox{$r_{\pi^{\pm} \pi^{\pm} }\  > 
\ r_{K^{\pm} K^{\pm}}\ > \ r_{\Lambda \Lambda}$} hierarchy
is observed which seems to disagree with the basic string (LUND)
model expectation. An adequate description of $r(m)$ is obtained
via the Heisenberg uncertainty relations and also by 
Local Parton Hadron Duality approach using a general QCD
potential. These lead to a relation of the type
$r(m) \simeq Constant/\sqrt{m}$. The present lack of knowledge 
on the $f_o(980)$ decay rate to the $K^0\overline{K}^0$ channel
prohibits the use of the $r_{K^0_SK^0_S}$ in the $r(m)$ analysis.
The use of a generalised BEC and I-spin invariance, which
predicts an BEC enhancement also in the $K^{\pm}K^0$ and 
$\pi^{\pm}\pi^0$ systems, should in the future help to 
include the $r_{K^0_SK^0_S}$ in the $r(m)$ analysis.
}
\section{Introduction}
\begin{figure}[t]
\hskip 0.5cm
\hspace*{0.5cm}\psfig{figure=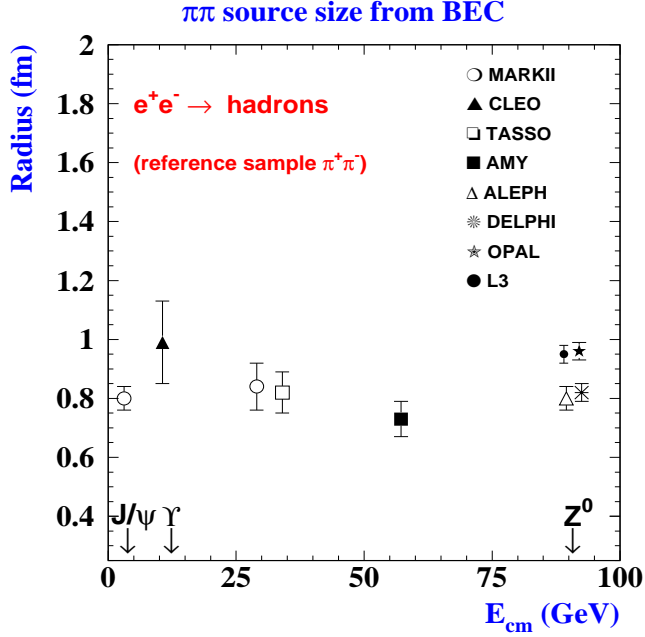,width=9.2cm,height=8.8cm}
\caption{A compilation of $r_{\pi^{\pm}\pi^{\pm}}$ measured in e$^+$e$^-$ annihilations.
 \label{fig:comp_pipi}}
\end{figure}

For more than three decades the Bose Einstein Correlations (BEC) of identical bosons,
mainly charged pions, were utilised to estimate the dimension of their emission
source. In nucleus-nucleus reactions a clear dependence of $r_{\pi\pi}$ 
on the size of the nucleus is observed. In $e^+ e^-$ annihilations leading to
hadronic final states no clear evidence is seen for a change in 
$r_{\pi\pi}$, the emission volume size,  
as the $\sqrt{s_{ee}}$ increases from 3.1 GeV (the $J/\psi$ mass)  to 
the Z$^0$ mass (see Fig. \ref{fig:comp_pipi}).
In particular if one remembers the large variety of experimental
procedure adopted by the different experiments. 
\noindent
Nevertheless, in order to be sure
to isolate the effect of the hadron mass we concentrated our study to 
$e^+ e^-$ annihilation at one energy only namely, at the Z$^0$ mass.  
At this energy the very large statistics of hadronic Z$^0$ decays of  
high average charged 
multiplicity, accumulated by the four LEP
experiments, ALEPH, DELPHI, L3 and OPAL, affords an excellent opportunity
to investigate whether the dimension $r$ of two-particle emitter size 
does depend on their mass.\\
In our analysis we rely on the 1-dimensional BEC studies where the emission source
is taken to be a sphere with a Gaussian density distribution. For the analysis 
the Lorentz invariant
variable $Q = \sqrt{-(q_1 - q_2)^2}$ is used together with the Goldhaber 
correlation parametrisation
$C(Q)= 1 + \lambda \exp(-r^2Q^2)$. Here $q_i$ are the 4-momenta of the
two identical bosons, $\lambda$ is the strength of the BEC effect and $r$ is the
emission volume size.\\
\section{Measurements of the source size ${\bf r}$}
\subsection{The $r_{\pi^{\pm}\pi^{\pm}}$ and $r_{K^{\pm}K^{\pm}}$} 
In Table 1 we present the weighted average value of 
the $r_{\pi^{\pm}\pi^{\pm}}$ measured in the Z$^0$ decays 
where the large systematic error reflects the spread of the individual $r$ values due to
the choices of the reference sample. In the same table 
are also listed the $r_{K^{\pm}K^{\pm}}$
derived by DELPHI from the
BEC of charged  Kaons and the preliminary value of
OPAL \cite{bec_kk} .
\subsection{The $r_{K^0_SK^0_S}$ and the Generalised BEC} 
The three LEP experiments, OPAL, DELPHI and ALEPH, have also measured the BEC of the
$K^0_SK^0_S$ pairs \cite{bec_k0k0} 
obtaining the $r_{K^0_SK^0_S}$ values listed 
in Table 1. The $K^0_SK^0_S$
pairs have three sources where two of them, the  $K^0K^0$ and 
$\overline{K}^0\overline{K}^0$,  
are identical di-bosons.
As for the boson-antiboson pair $K^0\overline{K}^0$, no BEC effect is expected if
one sums up all their possible decay pairs $K^0_SK^0_S, K^0_LK^0_S, K^0_SK^0_L$  
and $K^0_LK^0_L$. However by selecting only the $K^0_SK^0_S$ or the  $K^0_LK^0_L$ pairs
BEC enhancement is expected with a matching decrease of the $K^0_SK^0_L$ pairs.
Thus the $r_{K^0_SK^0_S}$ could have been used for our $r(m)$ investigation
was it not for the fact that the upper
side of the width of the enigmatic $f_o(980)$ resonance is above the $K\overline{K}$ 
threshold. 
Recently it has been shown \cite{isospin} that by invoking the generalised Bose-Einstein 
statistics and
using I-spin invariance one can evaluate the minimum BEC
contribution of $K^0_SK^0_S$ pairs
to the low mass enhancement. If the over-all
hadronic final state of the Z$^0$ decay is in a pure I=0, here denoted as
$\psi_o$, then one has the relation:\\
$\sum_X P[\psi_o \rightarrow K^0_S(\vec p) K^0_S(-\vec p) X] = 
(1/2) \cdot \sum_X P[\psi_o \rightarrow K^{\pm}(\vec p) K^{\pm}(-\vec p) X]
+ \sum_X P[\psi_o \rightarrow K^0 \overline{K}^0 X
\rightarrow K^0_S(\vec p) K^0_S(-\vec p)X]$ where X represents the hadrons accompanying the
Kaon pair.
From this relation follows that 
there will exist a BEC enhancement in the $K^0_SK^0_S$ pairs the height of which
will at least be half of that present 
in the charged identical Kaon pairs. Since this analysis has not been done so
far, the current quoted $r_{K^0_SK^0_S}$ values cannot serve our analysis.
Here is to note that the generalised Bose statistics predicts that 
BEC enhancement should also occur in the $\pi^{\pm}\pi^0$ 
(not studied so far) system in a similar strength to that 
seen in the $\pi^{\pm}\pi^{\pm}$ pairs.\\
The $r$ values obtained from the $\pi^{\pm}\pi^{\pm}$ and  $K^{\pm}K^{\pm}$ 
BEC analyses of  
DELPHI and the average values obtained from existing LEP 
experiments are shown in Fig 3. A first indication for the decrease of $r$ with the
increase of the hadron mass is seen but the data within the errors are still compatible
with a constant $r$ value, or even with a slight rise, as $m$ increases.
This situation calls for 
a BEC analysis of higher mass hadrons like e.g. the $\eta'(958)$ meson. However with  
its very low production rate of about 0.14 per Z$^0$ decay,
this avenue is closed to us. 
\renewcommand{\arraystretch}{1.6}
\begin{table}[t]
\caption{BEC results obtained at LEP1. (a) From spin composition analyses where 
no need exists for reference sample and Coulomb correction.  
\label{tab:comp_lep}}
 \begin{center}
    \begin{tabular}{|c|l|l|l|}
      \hline
        & ~~~~~~~~~~{ \bf {\large  \boldmath $\lambda$ }}  &~~~~~~~~~~{\bf r (fm)} & ~~{\bf  Experiment} \\
  \hline 
 { \large \boldmath  $\pi^{\pm} \pi^{\pm}$}  &  ~~~~~~~~~~$-$    & {$\bf 0.78 \pm 0.01 \pm 0.16$} & {\bf LEP Average}\cr
\hline
  ${ \boldmath \bf K^{\pm} K^{\pm}}$ & ${\bf 0.82 \pm 0.22 ~~^{+0.17}_{-0.12}}$ & ${ \bf 0.56 \pm 0.08 ~~^{+0.07}_{-0.06}}$  &{\bf OPAL(99)}  \\
       & ${\bf 0.82 \pm 0.11\pm0.25}$ & ${ \bf 0.48 \pm 0.04 \pm 0.07 }$ & {\bf DELPHI(96)} \\
 \hline
  ${\boldmath \bf K^0_S K^0_S}$ & ${\bf 1.14 \pm 0.23 \pm 0.32}$ & ${ \bf 0.76 \pm 0.10 \pm 0.11}$ & {\bf OPAL(95)}  \\
       & ${\bf 0.96 \pm 0.21 \pm 0.40}$ & ${ \bf 0.65 \pm 0.07 \pm 0.15}$ & {\bf ALEPH(94)}\\
       &  ${\bf 0.61 \pm 0.16 \pm 0.16}$ &  ${ \bf 0.55 \pm 0.08 \pm 0.12}$ & {\bf DELPHI(96)}  \\
\hline
 ${\boldmath \bf \Lambda\Lambda}~~~$  &  ~~~~~~~~~~$-$    & ${ \bf 0.11 \pm 0.02 \pm 0.01 }$ & {\bf ALEPH(99)} \\ 
\hline
 ${ \boldmath \bf \Lambda\Lambda}^{\boldmath \bf (a)}$ & ~~~~~~~~~~$-$ &${ \bf 0.19 ~~^{+0.37}_{-0.07}\pm 0.02 }$ & {\bf OPAL(96)}  \\
       &  ~~~~~~~~~~$-$    & ${ \bf 0.11 ~~^{+0.05}_{-0.03}\pm 0.01 }$ & {\bf DELPHI(98)} \\
       &  ~~~~~~~~~~$-$    & ${ \bf 0.17 \pm 0.13 \pm 0.04 }$ & {\bf ALEPH(99)} \\
\hline
\end{tabular}
\end{center}
\end{table}
\begin{figure}[t]
\centerline{\hbox{
\psfig{figure=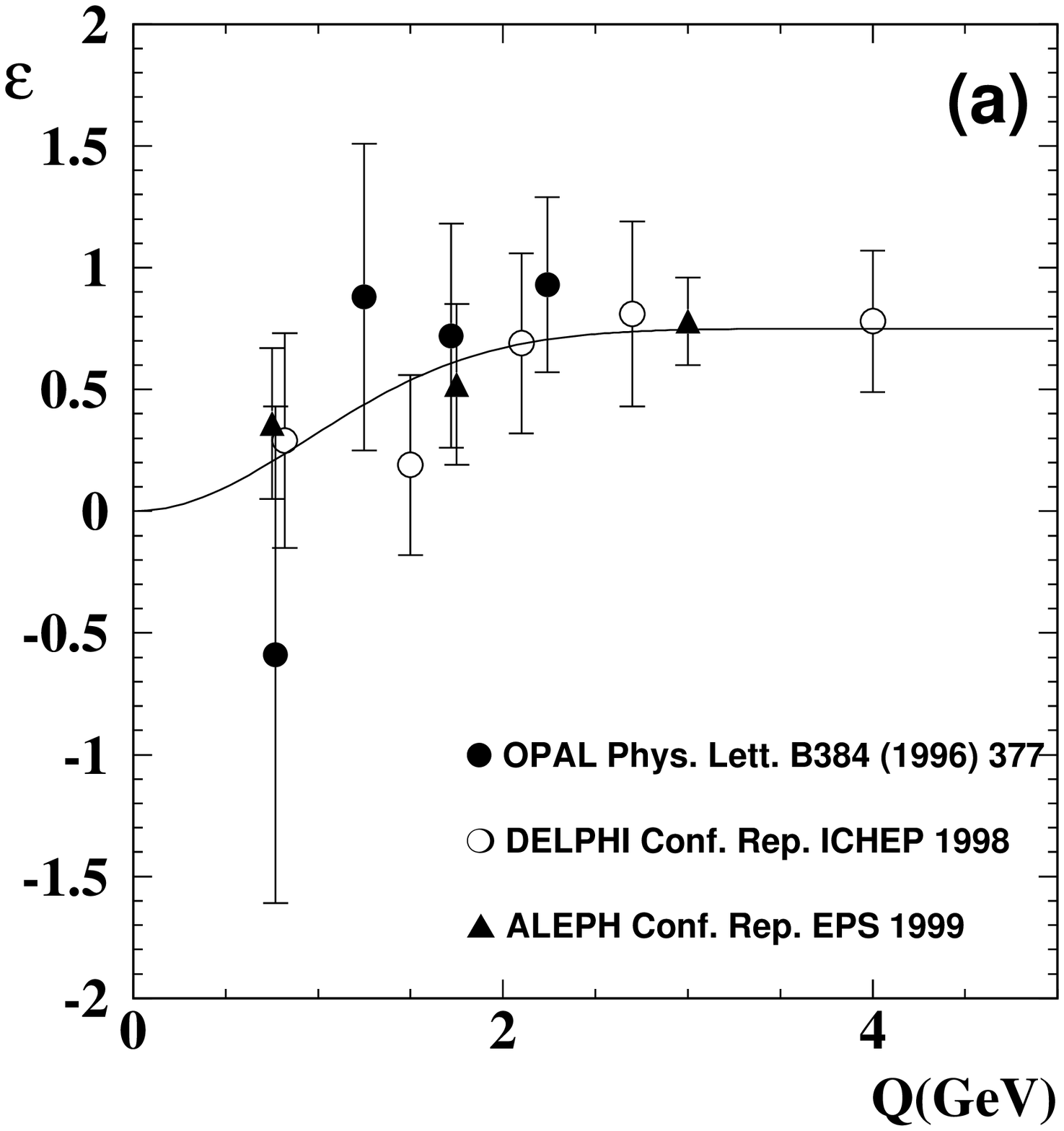,width= 6.9cm,height=7.3cm}
\hspace*{-0.2cm}\psfig{figure=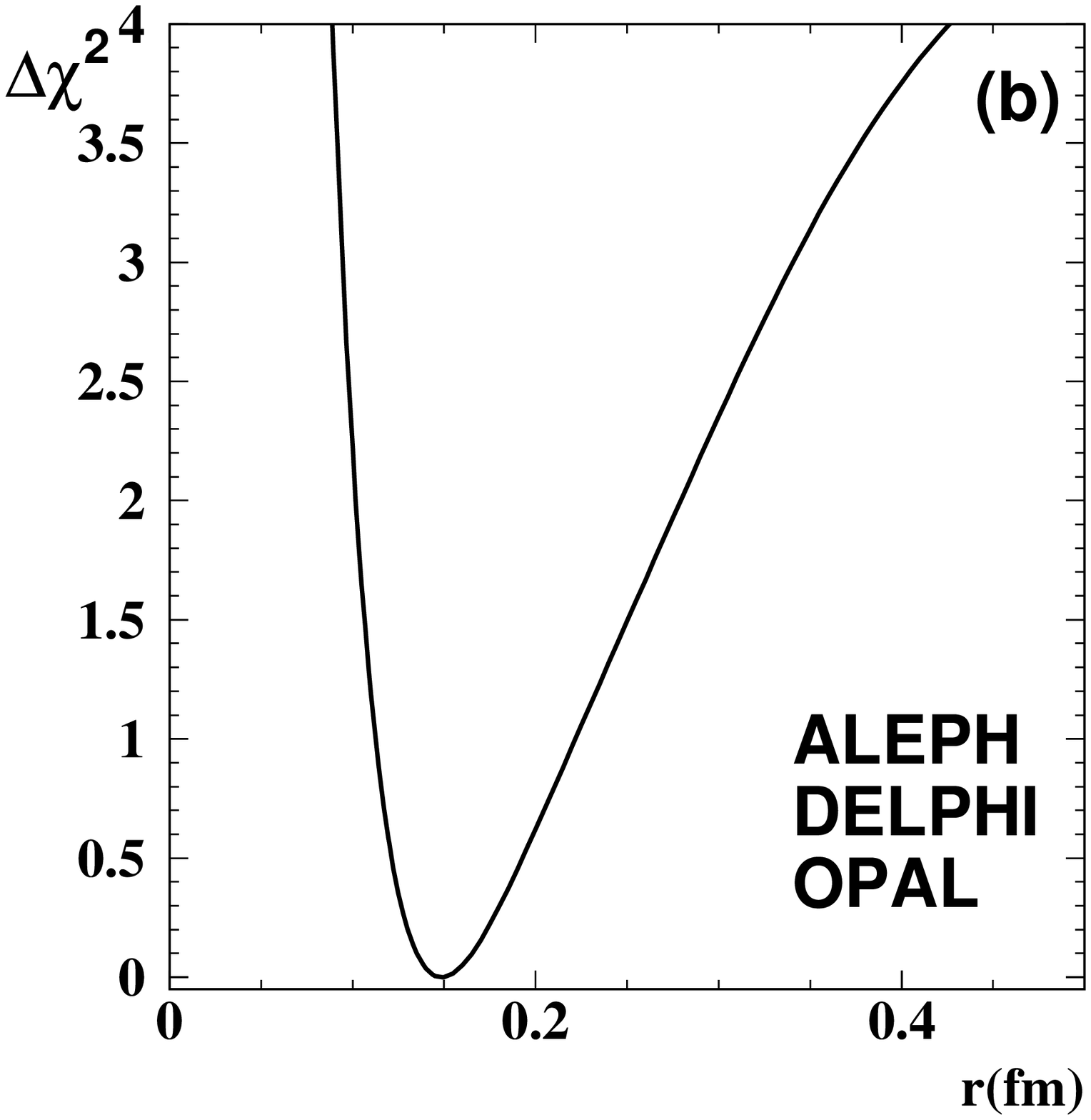,width=6.9cm,height=7.3cm}
}}
\caption{(a)
The S = 1 fraction, $\varepsilon$, of the
$\Lambda\Lambda(\bar \Lambda \bar \Lambda)$ pairs
measured as a function of $Q$ by the LEP collaborations.
The solid line represents the fit results as described in the text. 
(b) The  $\Delta\chi^2 = \chi^2 - \chi^2_{min}$ dependence on
$r_{\Lambda\Lambda}$ obtained in the fit to the combined LEP data. 
\label{fig:lamlam}}
\end{figure}
\subsection{Estimate of $r_{\Lambda\Lambda}$} 
\begin{figure}[t]
\begin{center}
\hspace*{0.5cm}\psfig{figure=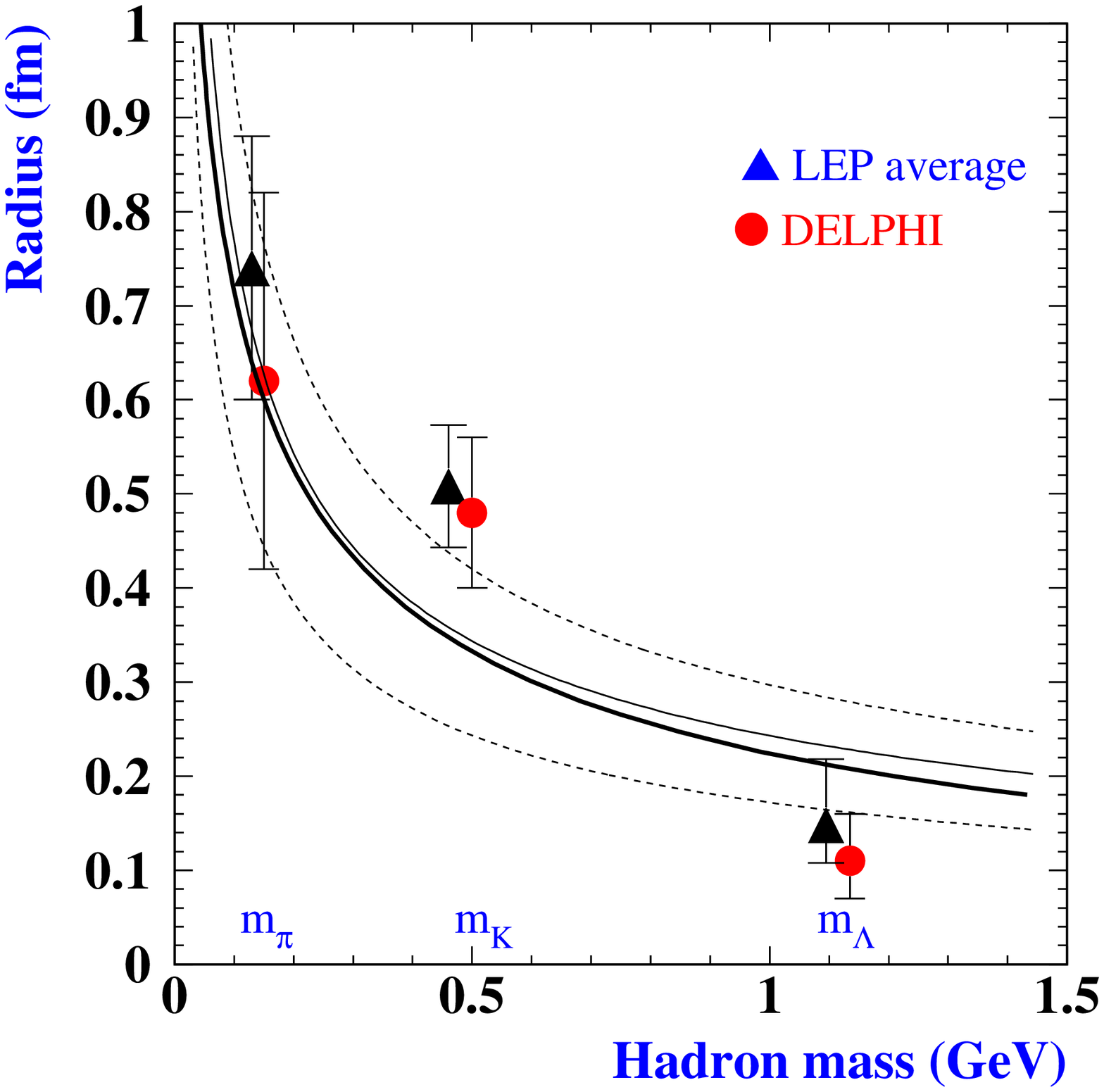,height=9.0cm,width=9.0cm}
\caption{The measured emitter radius $r$ as a function of the hadron
mass $m$ compared to some theoretical predictions (see text). \label{fig:main}}
\end{center}
\end{figure}

It was pointed out some time ago by G. Alexander and H.J. Lipkin 
\cite{alexlipkin}
that $r$ values for pairs of baryons can be measured by observing the onset of the 
Pauli exclusion
principle as the two identical baryons approach their threshold in their
centre of mass system. The ratio {\large $\varepsilon$}, defined as 
\mbox{{\large $\varepsilon$}$ = (S=1)/[(S=1) + (S=0)]$}, measures the
relative contribution of the S=1 state to the two $\Lambda$ system. Experimentally
this can be measured through the angular distribution  
of the angle between the decay protons of the two identical hyperons defined as 
$y^{\ast} = \cos({\bf p_1 p_2})$ . Here ${\bf p_1}$ and ${\bf p_2}$
are the momenta of the protons coming from the two $\Lambda$'s decay defined after two
transformations: first the $\Lambda$'s are transformed to their centre of mass system and
in the second each proton is transformed to the centre of mass of its parent 
$\Lambda$. Using this method OPAL, DELPHI and ALEPH have
estimated the $r_{\Lambda\Lambda}$ value which are listed in Table 1. In order to
determine the combined  $r_{\Lambda\Lambda}$ value of these three experiments we have
considered the values of {\large $\varepsilon$}(Q) shown in Fig. \ref{fig:lamlam}a. 
These values were fitted to the
expression {\large  $\varepsilon$}(Q)\ = \ 0.75 $[1 - e^{(-r^2_{\Lambda\Lambda}Q^2)}]$ 
with the result shown by the solid line in Fig. 2a corresponding to 
$$r_{\Lambda\Lambda} \ = \ 0.15 ^{+0.07}_{-0.04}\  fm.$$
\noindent
In Fig. 2b we show the distribution $\Delta\chi^2 \ = \ \chi^2- \chi^2_{min}$ 
with $\chi^2_{min}$ = 2.9.\\
The DELPHI values (circles) and the LEP averaged values (triangles) of $r_{\pi^{\pm}\pi^{\pm}}$, 
$r_{K^{\pm}K^{\pm}}$ and $r_{\Lambda\Lambda}$ are plotted in Fig. \ref{fig:main}  where a clear
hierarchy is seen, namely
 $$r_{\pi^{\pm}\pi^{\pm}}\ > \ r_{K^{\pm}K^{\pm}}\ > \
r_{\Lambda\Lambda}.$$ 
\noindent
The error bars represent the statistical and systematic
errors added in quadrature. The ALEPH collaboration has also attempted to
measure $r_{\Lambda\Lambda}$ by considering, similar to the BEC studies, the
ratio between the frequency of data $\Lambda$ pairs to that of 
Monte Carlo generated sample as a function of $Q$. The observed
depletion of pairs at low $Q$ was used to estimate $r_{\Lambda\Lambda} =
0.11 \pm 0.02 \pm 0.01$ fm (see Table 1). This method, which has a relative small
statistical error in comparison to the spin composition measurement, 
rests however on the assumption that the depletion of events
at low $Q$ is indeed due to the Pauli exclusion principle and not for any other reason.
This method, furthermore, has to rely  on a reference sample.\\ 
\section{Comparison between ${\bf r(m)}$ and theoretical models}
The fact that $r(m)$ decreases with the mass of the particle
can be checked against the predictions of multi-particle production models of 
$e^+ e^-$ annihilations. In particular this $r(m)$ behaviour seems to contradict 
the expectation of the string (LUND) model where $r(m)$ should increase 
with the particle mass.\\ 
On the other hand the $r(m)$ behaviour can be described in terms of the Heisenberg
uncertainty relations \cite{ourpaper} i.e., 
$\Delta p \Delta r \ = \ 2\ \mu \ v \ r \ = 
\ m\ v\ r \ = \ \hbar c$\ and \ 
$\Delta E \Delta t \ = 
2[p^2/(2m)]\Delta t \ = \ \hbar$ where $m$ is the mass of the hadron, $v$ and $p$ its 
velocity and momentum  
and $r$ is spatial separation between the two hadrons. In $\Delta E$, possible 
contributions from a potential energy are neglected.
From these equations one obtains
$$r(m) \ = \ c \sqrt{\hbar \Delta t}/\sqrt{m}\ \simeq \  0.243/\sqrt{m(GeV)}\ \ fm$$
when considering only the kinetic energy, equating $\Delta r = r$ and  
setting $\Delta t$ to $10^{-24}$ seconds 
representing the strong interaction time scale. 
As can be seen in Fig. \ref{fig:main} the thin solid line, 
which represents the expectation from the uncertainty  
relations, describes fairly well the data both in magnitude and in shape.
To illustrate the sensitivity to our particular choice of $\Delta t$ we also show
by the dashed lines the $r(m)$ expectations for $\Delta t = 1.5\times 10^{-24}$
and $0.5\times 10^{-24}$ seconds. Rearranging the expression for $r(m)$ 
it is amusing to find out that
$mr^2(m) =$ Const. $\simeq$ 0.06 \ GeV\ fm$^2$ i.e., 
the identical hadron pairs seem to emerge with the same moment of inertia 
irrespective of their mass value.\\ 
Another approach to describe the $r(m)$ dependence has been tried out\cite{ourpaper}. 
Namely,
by using the virial theorem in the frame work of 
the Local Hadron Parton Duality hypothesis. In this context
the general QCD potential given by
\vspace*{-1mm}
$$V(r)\,\,\,=\,\,\,\kappa\,\,r\,\,\,-\,\,\,\frac{4}{3}\,\frac{\alpha_S
\hbar c}{r}\,\,$$ was used
with the parameters set
of $\kappa = 0.14$ \ GeV$^2$ = 0.70 GeV/fm\  with
$\alpha_s \,\,= 2 \pi /9 \,\ln(\delta + \gamma/r )$, $\delta = 2 $ and
$ \gamma = 1.87$ \ GeV$^{-1} = 0.37$ fm as obtained from the hadron wave functions and 
decay constants. The result of this QCD based approach is shown in Fig. 3 by the 
solid thick line differs only very slightly from the uncertainty relations outcome. 
\section{Conclusion and summary}
The source dimension $r$ is found to change from $\approx$ 0.75 fm for 
charged pions to $\approx$ 0.15 fm for $\Lambda$'s. This trend is opposite 
in direction to
that expected in the basic string (LUND) model where $\partial r(m)/\partial m > 0$.
It can however be described in terms 
of the Heisenberg uncertainty relations as well as in the frame work of the
Local Parton Hadron Duality assumption. Both expect that 
$r(m) \simeq Constant/ \sqrt{m}$. This relation leads to the conclusion that
pairs of hadrons are produced with the same moment of inertia irrespective
of their mass value, which presents a challenge for future attempts to formulate 
multi-particle production models.  
More precise $r$ measurements for the source dimension
will be beneficial for the comparison with models. Of particular interest
will be the result from a BEC analysis of the $\eta \eta$ system which has a mass near
the K-meson but it is strange-less and its production rate in 
the Z$^0$ hadronic decays is relatively high. At the higher mass end there is little chance
to further explore the $r(m)$ behaviour. The production rate of the $\eta'(958)$
is too low and so are also the rates for the hyperons above the $\Lambda$ baryon.\\
Presently one cannot utilise for the $r(m)$ analysis the measured $r_{K^0_SK^0_S}$
values due to the fact that they may be influenced by the enigmatic $f_o(985)$ resonance decay
into the $K^0\overline{K}^0$ channel. In the frame work of the generalised BEC and 
I-spin invariance there is a possibility to separate the two
sources of the $K^0_SK^0_S$ system. This generalised BEC extension can be verified 
by observing BEC in the $\pi^{\pm}\pi^0$ pairs coming e.g. from hadronic Z$^0$ decays.

\section*{Acknowledgements}
Our thanks are due to H.J. Lipkin and E. Levin for many helpful discussions.

\end{document}